\documentclass[]{raa}
\usepackage{graphicx,times}
\usepackage{natbib}

\providecommand{\bm}[1]{\mbox{\boldmath$#1$\unboldmath}}

\begin{document}

\title{Modeling the Newtonian Dynamics for Rotation Curve Analysis 
   of Thin-Disk Galaxies}

 \volnopage{ {\bf 20xx} Vol.\ {\bf 9} No. {\bf XX}, 000--000}
   \setcounter{page}{1}

\author{James Q.~Feng and C. F. Gallo}
  
\institute{Superconix Inc,
      2440 Lisbon Avenue, 
      Lake Elmo, MN 55042, USA;
{\it info@superconix.com}\\
\vs \no
   {\small Received [year] [month] [day]; accepted [year] [month] [day] }
}

\abstract{
We present an efficient, robust computational method for 
modeling the Newtonian dynamics 
for rotation curve analysis of thin-disk galaxies.
With appropriate mathematical treatments, 
the apparent numerical difficulties associated with singularities 
in computing elliptic integrals 
are completely removed.
Using a boundary element discretization procedure,
the governing equations are transformed into a linear algebra matrix 
equation that can be solved by straightforward Gauss elimination
in one step without further iterations.
The numerical code implemented according to our algorithm
can accurately determine the surface mass density distribution
in a disk galaxy from a measured rotation curve (or vice versa).
For a disk galaxy with a typical flat rotation curve,
our modeling results show that the surface mass density
monotonically decreases from the galactic center toward periphery,
according to Newtonian dynamics.
In a large portion of the galaxy, the surface mass density follows
an approximately exponential law of decay with respect to
the galactic radial coordinate.
Yet the radial scale length for the surface mass density seems to be 
generally larger than that of the measured brightness distribution,
suggesting an increasing mass-to-light ratio 
with the radial distance in a disk galaxy.
In a nondimensionalized form, our mathematical system contains
a dimensionless parameter which
we call the ``galactic rotation number'' that represents
the gross ratio of centrifugal force and gravitational force.
The value of this galactic rotation number is determined as
part of the numerical solution.
Through a systematic computational analysis, 
we have illustrated that the galactic rotation number remains within
$\pm 10\%$ of $1.70$ for a wide variety of rotation curves.
This implies that the total mass in a disk galaxy 
is proportional to $V_0^2\,R_g$, 
with $V_0$ denoting the characteristic rotation velocity 
(such as the ``flat'' value in a typical rotation curve) and 
$R_g$ the radius of the galactic disk. 
The predicted total galactic mass of the Milky Way 
is in good agreement with
the star-count data.
}

   \authorrunning{J.~Q. Feng \& C.~F. Gallo}    %author_head in even pages
   \titlerunning{Modeling Newtonian Dynamics of Thin-Disk Galaxies}
                            % title_head in odd pages

\maketitle

\section{Introduction}

Observations have shown that
a galaxy is a stellar system consisting of a massive 
gravitationally bound assembly of 
stars, an interstellar medium of gas and cosmic dust, etc.
Many (mature spiral) galaxies share a common structure with the
{\em visible} matter distributed in a flat thin disk,
rotating about their center of mass 
in nearly circular orbits.
The behavior of the stellar systems such as galaxies is 
believed to be determined by Newton's laws of motion 
and Newton's law of gravitation \citep{binney87}. 
Thus, modeling the Newtonian dynamics of thin-disk galaxies 
is of fundamental importance to 
our understanding of the so-called ``galaxy rotation problem''--an
apparent discrepancy between the observed rotation of galaxies and
the predictions of Newtonian dynamics as generally perceived 
in the community of 
astrophysics \citep[e.g.,][]{freeman06, rubin06, rubin07, bennett07}.

Although scientifically well-established, 
the actual modeling of Newtonian dynamics, when applied to 
thin-disk galaxies, appeared in various forms in the literature
with inconsistent conclusions.
Without rigorous justification, 
some authors \citep[e.g.,][]{rubin06, rubin07, bennett07, sparke07, keel07}
tempted for simplicity to
apply formulas based on Keplerian dynamics to the thin-disk galaxies.
Theoretically, Keplerian dynamics can be derived from Newtonian dynamics 
as a special case
for spherically symmetric gravitational systems such as the solar system 
and, therefore, is not expected to provide accurate 
descriptions for thin-disk galaxies. 
Hence, serious efforts were made for integrating 
the Poisson equation with mass sources distributed on a disk, 
as summarized by \cite{binney87}.
The solution directly obtained from such efforts 
is the gravitational potential
which can yield the gravitational force 
by taking its gradient.
In an axisymmetric disk rotating at steady state, the gravitational force
(the radial gradient of gravitational potential)
is expected to equate to the centrifugal force due to rotation at
every point.

However, solving the disk-potential problem does not 
seem to be a trivial pursuit.
Traditional methods involved either 
treating the disk as a flattened spheroid that consists of 
a serious of thin homoeoids each having a uniform density 
\citep[e.g.,][]{brandt60, mestel63, cuddeford93}
or using the summation of 
modified Bessel functions for the potential 
\citep[e.g.,][]{toomre63,freeman70, nordsieck73, cuddeford93, conway00}.
Although seemingly elegant when derived in analytical formulas, 
those methods could yield closed-form solutions only for
a few special cases \citep[e.g.,][]{mestel63, freeman70, binney87}.
But for determining the mass distribution in a galactic disk 
from the measured rotation curve that could have 
a variety of shapes,
numerical integrations must be carried out and practical difficulties 
arise when those traditional analytical formulas are used.
For example, the flattened spheroid approach via Abel integral and 
its inversion intrinsically restricts the ``vertical'' mass distribution in the 
disk's axial direction to that dictated by the homoeoid structure rather than
that from observations \citep[e.g., according to]
[the scale heights of galactic disks
are nearly independent of radius]{vanderkruit82}.  
It is rather cumbersome to compute the surface mass density by 
integrating the mass density in spheroidal shells and 
the ``spheroid'' methods often lead to erroneous results for 
angular momentum analysis \cite[cf.][]{toomre63, nordsieck73}. 
The Bessel function approach leads to an integral extending to infinity,
whereas the observed rotation curve always ends at a finite distance.
Thus, it becomes necessary to construct orbital velocity beyond 
the observation limit based on various assumptions 
\citep[e.g.,][]{nordsieck73, bosma78, jalocha08}.
Moreover, the derivative of rotation velocity usually appearing
in the Bessel function formulation for computing mass density
tends to introduce significant errors in practical applications.  

In general, the fundamental solution to 
the Poisson equation (that governs the gravitation potential)
is called Green's function \citep{arfken85, cohl99}.
The potential from arbitrarily distributed sources can be
obtained by integrating the Green's function--serving 
as the integral kernel--multiplied by the source
density throughout the region where the sources are located.
Thus, considering the gravitational potential in terms of Green's function 
is the most direct approach for 
realistic modeling the galactic rotation dynamics
\citep[e.g.,][]{eckhardt02, pierens04, hure05}.
For sources distributed axisymmetrically on a thin disk,
the Green's function can be expressed in terms of the 
complete elliptic integral of the first kind \citep[e.g.,][]{binney87}.
Because the dynamics of thin-disk galactic rotation is 
typically described along the midplane ($z = 0$) with the mass distribution
being symmetric about the disk midplane and about its central axis, 
the radial gradient of potential in the midplane must be evaluated.
The elliptic integrals of the first kind and second kind that appear
in the radial gradient of potential can become mathematically singular 
at the midplane (when $z = 0$) where the radius of the source approaches 
that of the observation point.
Such singularities have been considered 
``inconvenient from the point of view of numerical work''
by \cite{binney87} 
and ``bothersome'' by \cite{eckhardt02}.
Methods were suggested to circumvent such singularities 
by evaluating the radial gradient of potential at 
a vertical distance $z$ slightly away from $z = 0$
\citep[cf][]{binney87, eckhardt02}, which 
seem to be somewhat {\em ad hoc} by nature and 
lack of desirable mathematical elegance. 
On the other hand, it is the axisymmetric mass distribution
within an idealized rotating infinitesmally thin disk 
that has often been of practical interest 
especially for rotation curve analysis
\citep{toomre63}.
Therefore, the efforts of effectively dealing with
the singularities arising from elliptic integrals has been continuously
made for robust and accurate computations of 
the disk galaxy rotation problem 
\citep[even up to recent years, e.g.,][]{eckhardt02, pierens04, 
hure05, hure09}.

In the present work, 
we derive a numerical model for computing the Newtonian dynamics
of thin-disk galactic rotation that allows the mass to be distributed even 
in an 
infinitesmally thin region around the midplane of the disk 
with the governing equation being considered strictly 
along the midplane ($z = 0$) and 
the singularities from elliptic integrals 
treated rigorously based on the concept of the mathematical limit.  
To enable dealing with arbitrary forms of rotation curves and 
mass density distributions, 
we adopt the techniques developed with boundary element method 
\citep[cf.][]{sladek98, gray98,sutradhar08} for solving integral equations 
using compactly supported basis functions
instead of that extending to infinity like Bessel functions,
as detailed in \S~2.
Hence the finite physical problem domain for disks of finite sizes 
can be conveniently 
considered, without the need of
speculated rotation curve beyond the ``cut-off'' radius 
and evaluation of 
the derivative of rotation velocity.
By nondimensionalizing the governing equations,
a dimensionless parameter 
which we call the ``galactic rotation number'' appears in
the force balance (or centrifugal-equilibrium) equation,
representing the gross ratio of centrifugal force and gravitational force.
We show that together with a constraint equation for mass conservation,
the value of this galactic rotation number can be determined as part of 
the numerical solution, with computational examples 
presented in \S~3.
With a known value of the galactic rotation number, 
the total galactic mass can be determined from 
measured galactic radius and characteristic rotation velocity,
as shown in \S~4 wherefrom important 
physical insights are discussed.

\section{Mathematical Formulation and Computational Techniques}

\begin{figure}%[htb]
\resizebox{!}{0.56\textwidth}
{\includegraphics[clip=true,scale=0.48,angle=90,angle=90,angle=90,viewport=30 60 400 800]{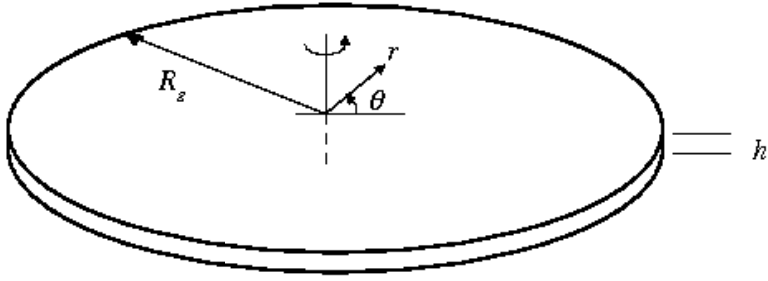}}
%{\includegraphics[clip=true,scale=0.24,viewport=100 116 400 700]{sketch}}
\caption{\label{fig:fig1}Definition sketch of the thin-disk model
considered in the present work.
The mass is assumed to distribute axisymmetrically in the circular disk
of uniform thickness $h$
with a variable density as a function of radial coordinate $r$
(but independent of the polar angle $\theta$).}
\end{figure}

For convenience of mathematical treatment,
we represent a rotating galaxy by 
a self-gravitating continuum of axisymmetrically distributed
mass in a circular disk with an edge at finite radius $R_g$,
as shown in Fig.~\ref{fig:fig1}.
This kind of continuum representation is typically valid 
when the distributed masses are viewed on a scale that is 
small compare to the size of the galaxy,
but large compared to the mean distance between stars. 
Without loss of generality, we consider 
the thin disk having a
uniform thickness ($h$) with a variable mass density ($\rho$)
as a function of radial coordinate ($r$).
Because we consider the situation of thin disk, 
the vertical distribution of mass (in the $z$-direction) is expected to
contribute inconsequential dynamical effect 
especially as the disk thickness becomes
infinitesmal. 
In mathematical terms, the meaningful variable here is
actually the surface mass density
$\sigma(r) \equiv \rho(r) \, h$. 
Whether to consider the surface mass density $\sigma(r)$ 
or the bulk mass density $\rho(r)$ in the mathematical equations 
is really a matter of taste,
since they can easily be converted to each other 
using a constant factor $h$ by our definition. 
In the present work, we use the bulk density $\rho(r)$ for 
its consistency with the direct physical 
perception of a thin disk with a nonzero thickness $h$.

Physically, the stars in a galaxy must rotate about the 
galactic center to maintain the disk-shape mass distribution.
Without the centrifugal effect due to rotation, 
the stars would collapse into the galactic center 
as a result of the gravitational field among themselves.
According to \cite{binney87},
it is also reasonable to assume the galaxy is 
in an approximately steady state with the gravitational force
and centrifugal force balancing each other,
in view of the fact that most disk stars have completed a large
number of revolutions.
    
\subsection{Governing Equations}
Instead of following the traditional approach by first
solving gravitational potential from the Poisson equation,
we derive the governing equation directly from the 
consideration of force balance.
Here, the force density on a test mass at 
the point of observation ($r$, $\theta = 0$)
generated by the gravitational attraction due to the 
summation (or integration) of a distributed mass density $\rho(\hat{r})$
at position described by the variables of integration ($\hat{r}$, 
$\hat{\theta}$) is expressed as an integral over the entire disk,
with the distance between ($r$, $\theta = 0$) and 
($\hat{r}$, $\hat{\theta}$) given by $(\hat{r}^2 + r^2 - 2 \hat{r} \, r \,
\cos \hat{\theta})^{1/2}$
and the vector projection given by $(\hat{r} \cos \hat{\theta} - r)$. 
Thus, the equation for gravitational force
to balance the centrifugal force at each and every point
in a thin disk can be written as (according to Newton's laws)
\begin{eqnarray} \label{eq:force-balance0}
\int_0^1 \left[\int_0^{2 \pi}
\frac{(\hat{r} \cos \hat{\theta} - r) d\hat{\theta}}
{(\hat{r}^2 + r^2 - 2 \hat{r} r \cos \hat{\theta})^{3/2}}\right]
\rho(\hat{r}) h \hat{r} d\hat{r} %\nonumber\\
+ A \frac{V(r)^2}{r}
 = 0 \, ,
\end{eqnarray}
where all the variables are made dimensionless
by measuring lengths (e.g., $r$, $\hat{r}$, $h$)
in units of the outermost galactic radius $R_g$,
disk mass density $\rho$ in units of
$M_g / R_g^3$ with $M_g$ denoting the total galactic mass,
and rotation velocities $V(r)$ in units of the
a characteristic galactic rotational velocity $V_0$
(usually defined according to problem of interest, e.g., such as 
the maximum velocity corresponding to the flat part of a rotation curve).
The disk thickness $h$ is assumed to be constant and small
in comparison with the galactic radius $R_g$.
Our results for surface mass density $\rho(r) \, h$
are expected to be insensitive to the exact value of this ratio
as long as it is small.
There is no difference in terms of physical meaning between
the notations $(r, \theta)$ and $(\hat{r}, \hat{\theta})$;
but mathematically the former denotes the independent variables in 
the integral equation (\ref{eq:force-balance0})
whereas the latter the variables of integration. 
The gravitational force represented as the summation of
a series of concentric rings is described
by the first (double integral) term while the centrifugal force
by the second term in 
(\ref{eq:force-balance0}).

Our process of nondimensionalization of the force-balance equation
yields a dimensionless parameter,
which we call the ``galactic rotation number'' $A$, as given by
\begin{equation} \label{eq:parameter-A}
A \equiv \frac{V_0^2 \, R_g}{M_g \, G} \, ,
\end{equation}
where $G$ ($= 6.67 \times 10^{-11}$ (m$^3$ kg$^{-1}$ s$^{-2}$))
denotes the gravitational constant,
$R_g$ is the outermost galactic radius,
and $V_0$ is the characteristic velocity
(which is equated here to the maximum asymptotic rotational velocity).
This galactic rotation number $A$
simply indicates the relative importance of
centrifugal force versus gravitational force.

Equation (\ref{eq:force-balance0}) can either be used to 
determine the surface mass density $\rho(r)\,h$ 
from a given rotation curve $V(r)$ or vice versa.
But when both $\rho(r)$ and $A$ are unknown, 
another independent equation is needed to have a well-posed 
mathematical problem.
According to the law of conservation of mass, 
the total mass of the galaxy $M_g$ should 
be constant satisfying the constraint
\begin{equation} \label{eq:mass-conservation}
2 \pi \int_0^1 \rho(\hat{r}) h \hat{r} d\hat{r} = 1.
\end{equation}
This constraint can be used for determining the value of 
galactic rotation number $A$
while (\ref{eq:force-balance0}) for $\rho(r)$.
Equations (\ref{eq:force-balance0})-(\ref{eq:mass-conservation})
can in principle be used to
determine the mass density distribution $\rho(r)$ in the disk,
the galactic rotation number $A$, and the total galactic mass $M_g$,
all from measured values of $V(r)$, $R_g$, $V_0$, and $h$.
On the other hand,
if $\rho(r)$ and $h$ (or $\rho(r)\,h$) are known, 
$V(r)$ can of course be determined from (\ref{eq:force-balance0}).

Moreover, it is known that
the integral with respect to $\hat{\theta}$ in (\ref{eq:force-balance0})
can be written as
\begin{eqnarray} \label{eq:elliptic-integral-form}
\int_0^{2 \pi}
\frac{(\hat{r} \cos \hat{\theta} - r) d\hat{\theta}}
{(\hat{r}^2 + r^2 - 2 \hat{r} r \cos \hat{\theta})^{3/2}} %\nonumber\\
%\quad \quad \quad \quad \quad \nonumber\\
= 2 \left[\frac{E(m)}{r (\hat{r} - r)} - \frac{K(m)}{r (\hat{r} + r)}\right]
 \, ,
\end{eqnarray}
where $K(m)$ and $E(m)$ denote the complete elliptic integrals
of the first kind and second kind,
with
\begin{equation} \label{eq:m-def}
m \equiv \frac{4 \hat{r} r}{(\hat{r} + r)^2} \, .
\end{equation}
Thus, (\ref{eq:force-balance0}) can be expressed in a simpler form 
\begin{equation} \label{eq:force-balance}
\int_0^1 \left[
\frac{E(m)}{\hat{r} - r} - \frac{K(m)}{\hat{r} + r}
\right]
\rho(\hat{r}) h \hat{r} d\hat{r}
+ \frac12 A V(r)^2
 = 0 \, ,
\end{equation}
which is more suitable for 
the boundary element type of numerical implementation
(with the double integral converted to a single integral). 

\subsection{Computational techniques}
Following a standard boundary element approach 
\citep[e.g.,][]{sladek98, gray98,sutradhar08},
the governing equations (\ref{eq:force-balance}) and
(\ref{eq:mass-conservation})
can be discretized by dividing the one-dimensional
problem domain $0 \le r \le 1$ into a finite number of line segments
called (linear) elements.
Each element covers a subdomain confined by two end nodes,
e.g., element $n$ corresponds to the subdomain
$[r_n, r_{n+1}]$, where $r_n$ and $r_{n+1}$ are nodal values of
$r$ at nodes $n$ and $n+1$, respectively.
On each element, which is mapped onto a unit line segment $[0, 1]$ in
the $\xi$-domain (i.e., the computational domain),
$\rho$ is expressed in terms of the linear basis functions as
\begin{equation} \label{eq:rho-xi}
\rho(\xi) = \rho_n (1 - \xi) + \rho_{n+1} \xi \, , \quad 0 \le \xi \le 1 \, ,
\end{equation}
where $\rho_n$ and $\rho_{n+1}$ are nodal values of $\rho$ at
nodes $n$ and $n + 1$, respectively.
Similarly, the radial coordinate $\hat{r}$
on each element is also expressed
in terms of the linear basis functions by
so-called isoparameteric mapping:
\begin{equation} \label{eq:r-xi}
\hat{r}(\xi) = \hat{r}_n (1 - \xi) + \hat{r}_{n+1} \xi \, , 
\quad 0 \le \xi \le 1 \, .
\end{equation}
If the rotation curve $V(r)$ is given (from observational measurements),
the $N$ nodal values of $\rho_n = \rho(r_n)$ are determined by
solving $N$ independent residual equations over $N - 1$ element
obtained from
the collocation procedure, i.e.,
\begin{eqnarray} \label{eq:force-balance-residual}
\sum_{n = 1}^{N - 1} \int_0^1 \left[
\frac{E(m_i)}{\hat{r}(\xi) - r_i} - \frac{K(m_i)}{\hat{r}(\xi) + r_i}
\right]
\rho(\xi) h \hat{r}(\xi) \frac{d\hat{r}}{d\xi} d\xi %\nonumber \\
+ \frac12 A V(r_i)^2
 = 0 \, , \quad i = 1, 2, ..., N \, ,
\end{eqnarray}
with
\begin{equation} \label{eq:mi-def}
m_i(\xi) \equiv \frac{4 \hat{r}(\xi) r_i}{[\hat{r}(\xi) + r_i]^2} \, ,
\end{equation}
where $\rho(\xi) = \rho_n(1 - \xi) + \rho_{n+1} \xi$. 
The value of $A$ can be solved by the addition of
the constraint equation
\begin{equation} \label{eq:mass-conservation-residual}
2 \pi \sum_{n = 1}^{N - 1} \int_0^1
\rho(\xi) h \hat{r}(\xi) \frac{d\hat{r}}{d\xi} d\xi - 1 = 0 \, .
\end{equation}
Thus, we have $N + 1$ independent equations for determining
$N + 1$ unknowns.
The mathematical problem is well-posed.
The set of linear equations comprising (\ref{eq:force-balance-residual})
and (\ref{eq:mass-conservation-residual})
for $N + 1$ unknowns (i.e., $N$ nodal values of $\rho_n$ and $A$),
once computed with appropriate treatments of the 
mathematical singularities shown in Appendix A,
can be transformed into a matrix form
using the Newton-Raphson method and then
solved with a standard matrix solver,
e.g., by Gauss elimination in one step without 
further iterations \citep{press88}.

\section{Computational Examples}

As we mentioned before, equations (\ref{eq:force-balance-residual})
and (\ref{eq:mass-conservation-residual}) can be used to
either solve for $\rho(r)$ and $A$ from a given rotation curve $V(r)$
or determine the rotation curve $V(r)$ from a given 
surface mass density distribution $\sigma(r) = \rho(r)\, h$.
Usually, solving for $\rho(r)$ from a given rotation curve $V(r)$
requires computation of a linear algebra matrix problem 
whereas determining $V(r)$ from a given $\rho(r)$ only involves 
a straightforward integration.
But in a spiral galaxy it is the rotation curve that can be 
measured with considerable accuracy; 
therefore, the observed rotation curve has been regarded to 
provide the most reliable means for determining
the distribution of gravitating matter 
therein \citep{toomre63, sofue01}.
Hence, we first consider examples of solving for $\rho(r)$ and $A$ 
from a given $V(r)$.

\subsection{Mass distribution for rotation curve of typical shape}

To obtain numerical solutions,
the value of (constant) disk thickness $h$ must be provided;
we assume $h = 0.01$, 
which is typical of disk galaxies like the Milky Way. 
For computational efficiency, 
we distribute more nodes in the regions 
(e.g., near the galactic center and disk edge)
where $\rho$ has a greater
gradient of variations.
The typical number of nonuniformly distributed nodes $N$ 
used in the computation
is $1001$ with which we found for most cases to be sufficient for
obtaining a smooth curve of $\rho$ versus $r$ and 
discretization-insensitive values of galactic rotation number $A$.
When numerically integrating element-by-element in 
(\ref{eq:force-balance-residual})
and (\ref{eq:mass-conservation-residual}),
we use ordinary 6-point Gausian quadrature for 
integrals with respect to 
$0 \le \xi \le 1$.
The two-dimensional integrals (\ref{eq:log-integral-identities})
on a singular element are calculated 
numerically by ordinary $6 \times 6$-point 
Gausian quadrature on a unit square with 
$0 \le \eta \le 1$ and
$0 \le \xi \le 1$.

\begin{figure}%[htb]
\resizebox{!}{0.72\textwidth}
{\includegraphics[clip=true,scale=0.82,viewport=66 330 760 760]{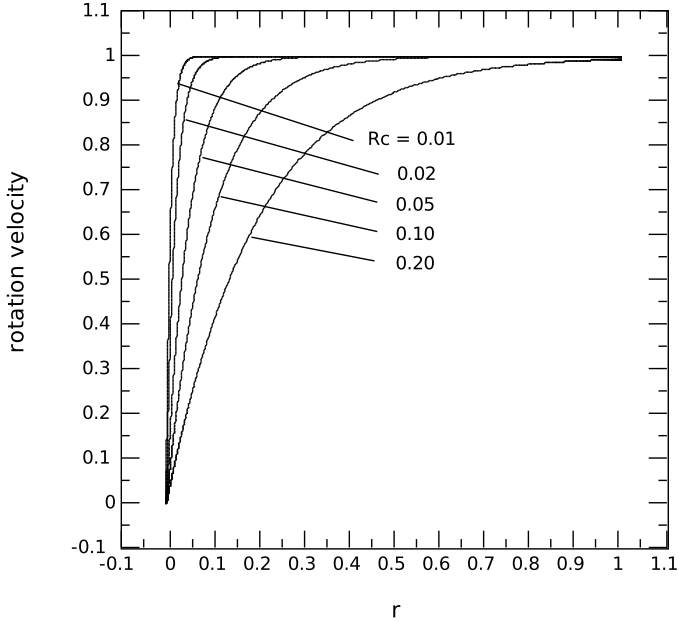}}
\caption{\label{fig:fig2} Nondimensionalized orbital velocity profiles
$V(r)$ according to mathematically idealized description (\ref{eq:V-ideal}) 
for $R_c = 0.01$, $0.02$, $0.05$, $0.1$, and $0.2$.}
\end{figure}

The measurements of galactic rotation curve 
of mature spiral galaxies reveal that 
the rotation velocity $V(r)$ typically rises linearly
from the galactic center in a small core and then
bend down to reach an approximately constant value extending 
to the galactic periphery \citep{rubin70,roberts75, bosma78,rubin80}.
These basic features may be mathematically idealized as
\begin{equation} \label{eq:V-ideal}
V(r) = 1 - e^{-r/R_c} 
\, ,
\end{equation}
where the dimensionless orbital velocity $V(r)$ is measured in units of
the characteristic velocity $V_0$ defined as the maximum orbital velocity,
and the parameter $R_c$ can serve as the scale of 
the ``core'' of a galaxy.
As shown in Fig~\ref{fig:fig2},
close to the galactic center when $r/R_c$ is small, 
we have $V(r) \sim r/R_c$ describing a linearly rising rotation velocity
(by virtue of the Taylor expansion of $e^{-r/R_c}$).  
The initial slope of this 
rising rotation velocity is given by $1/R_c$. 
Thus, larger value of $R_c$ leads to a more gradual rise of 
the rotation velocity and a shrinking ``flat'' part of rotation curve
which seems to disappear for $R_c \ge 0.2$.

\begin{figure}%[htb]
\resizebox{!}{0.72\textwidth}
{\includegraphics[clip=true,scale=0.82,viewport=66 330 760 760]{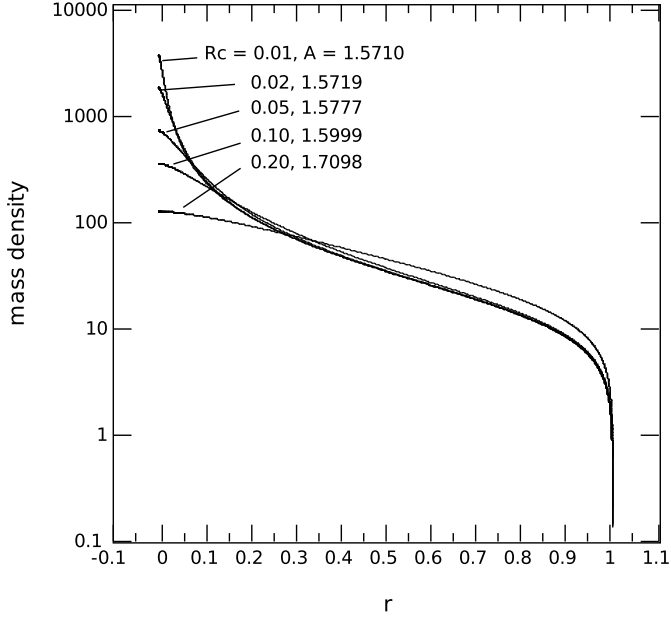}}
\caption{\label{fig:fig3}The distributions of
mass density $\rho(r)$ computed for
$R_c = 0.01$, $0.02$, $0.05$, $0.1$, and $0.2$
with $A = 1.5710$, $1.5719$, $1.5777$, $1.5999$, and $1.7098$
determined as part of the numerical solutions.}
\end{figure}

Corresponding to the rotation curves in Fig \ref{fig:fig2} 
as described by (\ref{eq:V-ideal}),
the computed mass density distributions in galactic disk
are shown in Fig.~\ref{fig:fig3}.
For $R_c \le 0.02$, the curves of $\rho$ versus $r$ 
approach an asymptotic one for the most part except in a 
tiny region around galactic center 
where the peak density value at $r = 0$ still increases 
with further decreasing $R_c$.
In other words,
the mass density tends to decrease rapidly from the galactic center
(with a slope becoming steeper for a tigher galactic core
with a smaller $R_c$).
However, beyond $r = R_c$, the mass density decrease more
gradually towards the galactic periphery 
until reaching the galactic edge where it takes a sharp drop.
Outside the galactic core ($r > R_c$),
only for $R_c > 0.1$ do changes in mass density 
distribution and the value of $A$
become noticeable with varying $R_c$.
Noteworthy here is that the computed values of
galactic rotation number $A$ for $R_c \le 0.15$
are within a small interval [$1.5708$, $1.6422$]
despite orders of magnitude of 
$R_c$ variation.
It appears that as $R_c \to 0$ the value of $A$ approaches 
a limit at $\sim1.5708$.  For example, the computed results show 
that $A = 1.57085$ and $1.57080$ for $R_c = 0.005$ and $0.001$, respectively.
But the increase of $A$ with $R_c$ becomes more significant 
for $R_c > 0.15$, as illustrated by the computed results at 
$R_c = 0.2$ and $0.3$
yielding $A = 1.7098$ and $1.9224$, respectively.

At the limit of $R_c \to 0$, the (idealized) rotation curve 
as described by (\ref{eq:V-ideal}) approaches 
a completely flat one $V(r) = 1$ (except in the 
infinitesmal neighborhood of $r = 0$).  
The solution at this limit should approach that 
of the well-known Mestel's disk \citep{mestel63} given by
\begin{equation} \label{eq:Mestel}
\rho(r) = \frac{A}{2 \pi h r}\left[1 - \frac{2}{\pi} \sin^{-1}(r)\right]
\, ,
\end{equation}
in a dimensionless form consistent with the nomenclature of 
the present work.
Here, according to (\ref{eq:mass-conservation})
the galactic rotation number $A$ can be determined by
\begin{equation} \label{eq:A-Mestel}
A = \frac{1}{\int_0^1\left[1 
- \frac{2}{\pi} \sin^{-1}(\hat{r})\right]\,d\hat{r}} = \frac{\pi}{2}
= 1.5707963
\, .
\end{equation}

As a test, we can substitute $\rho(r)$ given by (\ref{eq:Mestel}) 
into (\ref{eq:force-balance-residual}) 
and (\ref{eq:mass-conservation-residual})
and compute with our code for numerical integrations
to determine $V(r)$ and $A$.
With the first node at $r = 0$ being ignored to avoid 
the numerical difficulties with the singularity of $\rho$ in
(\ref{eq:Mestel}),
we can indeed obtain a flat $V(r) = 1$ throughout the entire interval $(0, 1]$
(except in an infinitesmal neighborhood around $r = 0$) and 
$A = 1.57081$.
The computed curve of $\rho$ versus $r$ corresponding to (\ref{eq:Mestel})
with $A = 1.57081$ 
overlaps that of $R_c = 0.01$ in Fig~\ref{fig:fig3}
(except in the infinitesmal neighborhood of $r = 0$), as expected. 
This exercise demonstrates our code capability for determing 
the rotation curve from a given disk mass distribution, and
also in a way verifies the correctness of 
our computational code implementation.
Since most Sb galaxies--intermediate type of spiral galaxies--have 
rotation curves typically with a very steep rise 
in a small central core region,
the mass density distribution in those Sb galaxies 
(including the Milky Way) is expected to be
reasonably approximated by that of the Mestel disk (\ref{eq:Mestel}).
But for less massive Sc galaxies having more gradual rise 
rotation curves, their mass density distributions can
deviate noticeably from that of the Mestel disk especially 
toward the galactic center, as shown in Fig~\ref{fig:fig3} 
for those with $R_c > 0.02$.

\subsection{Rotation curve for given mass distribution}

As demonstrated in \S~3.1, numerically computing the integration
in (\ref{eq:force-balance-residual})
for a given $\rho(\hat{r})$ as that of Mestel's disk
can produce a completely flat rotation curve.
Actually, rotation curves similar to those in Fig~\ref{fig:fig2}
can also be produced 
by a combination of the Freeman exponential disk and Mestel disk.
Here, the Freeman disk has a surface mass density 
proportional to $e^{-r/R_d}$ with $R_d$ denoting a scale length
for the exponental disk \citep{freeman70}.
But the Freeman exponential disk alone is known not to be able
to produce a rotation curve with considerable flat portion
as often being observed in disk galaxies
\citep[e.g.,][]{freeman70, binney87}.
The case of $V(r)$ for the Freeman exponential disk 
can also be computed with 
our code, as a check; the result showed excellent agreement with that 
of Freeman's analytical formula. 
If we use
the Freeman disk for describing the galactic core having a 
rising rotation velocity and Mestel disk for the outer flat part, 
there is a good chance to obtain rotation curves of 
typically observed shapes.
For example, we can simply construct a mass density model 
(which we call the Freeman-Mestel model) as 
\begin{eqnarray} \label{eq:freeman-mestel_rho}
\rho(r) = \left\{
\begin{array}{ccc}
\rho_0 \, e^{- r / R_d} \, , \qquad \qquad \qquad 0 \le r < \tilde{R}_c
\\ \\
\frac{A}{2 \pi h \, r}
\left[1 - \frac{2}{\pi} \sin^{-1}(r)\right] \, , \quad  
\tilde{R}_c \le r \le 1 \,
\end{array}
\right . \, ,
\end{eqnarray}
where
\[
R_d =
\left\{\frac{1}{\tilde{R}_c} + \frac{2}{\pi \sqrt{1 - \tilde{R}_c^2}
[1 - 2 \sin^{-1}(\tilde{R}_c) / \pi]}\right\}^{-1}
\]
and
\[
\rho_0 = \frac{A}{2 \pi h \, \tilde{R}_c e^{-\tilde{R}_c / R_d}}
\left[1 - \frac{2}{\pi} \sin^{-1}(\tilde{R}_c)\right]
\, ,
\]
so that both $\rho$ and $d\rho / dr$ are continuous at 
the connecting point $r = \tilde{R}_c$.
Moreover,
the mass conservation constraint (\ref{eq:mass-conservation-residual})
can be used to determine the value of galactic rotation number as
\begin{equation} \label{eq:freeman-mestel_A}
A = \left[2 \, \pi \, \sum_{n=1}^{N-1} \int_0^1 \rho^{*}(\xi) h \hat{r}(\xi)
\frac{d \hat{r}}{d\xi} d\xi \right]^{-1} \, ,
\end{equation}
where $\rho^{*}$ comes from that given by (\ref{eq:freeman-mestel_rho}) by
setting $A = 1$.

\begin{figure}%[htb]
\resizebox{!}{0.72\textwidth}
{\includegraphics[clip=true,scale=0.82,viewport=60 330 760 760]{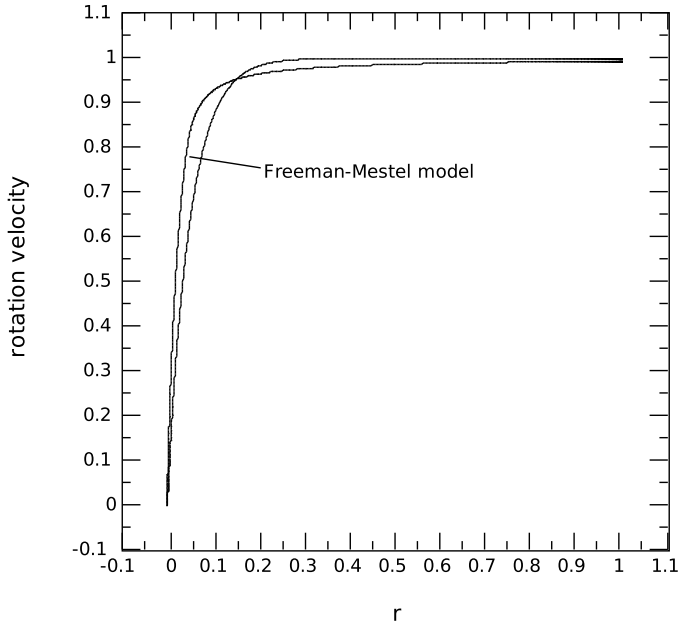}}
\caption{\label{fig:fig4}The rotation velocity $V(r)$ 
determined with $\rho(r)$ given by (25) for the 
Freeman-Mestel model at $\tilde{R}_c = 0.05$,
compared with that in Fig~\ref{fig:fig2} for
$R_c = 0.05$.}
\end{figure}

\begin{figure}%[htb]
\resizebox{!}{0.72\textwidth}
{\includegraphics[clip=true,scale=0.82,viewport=60 330 760 760]{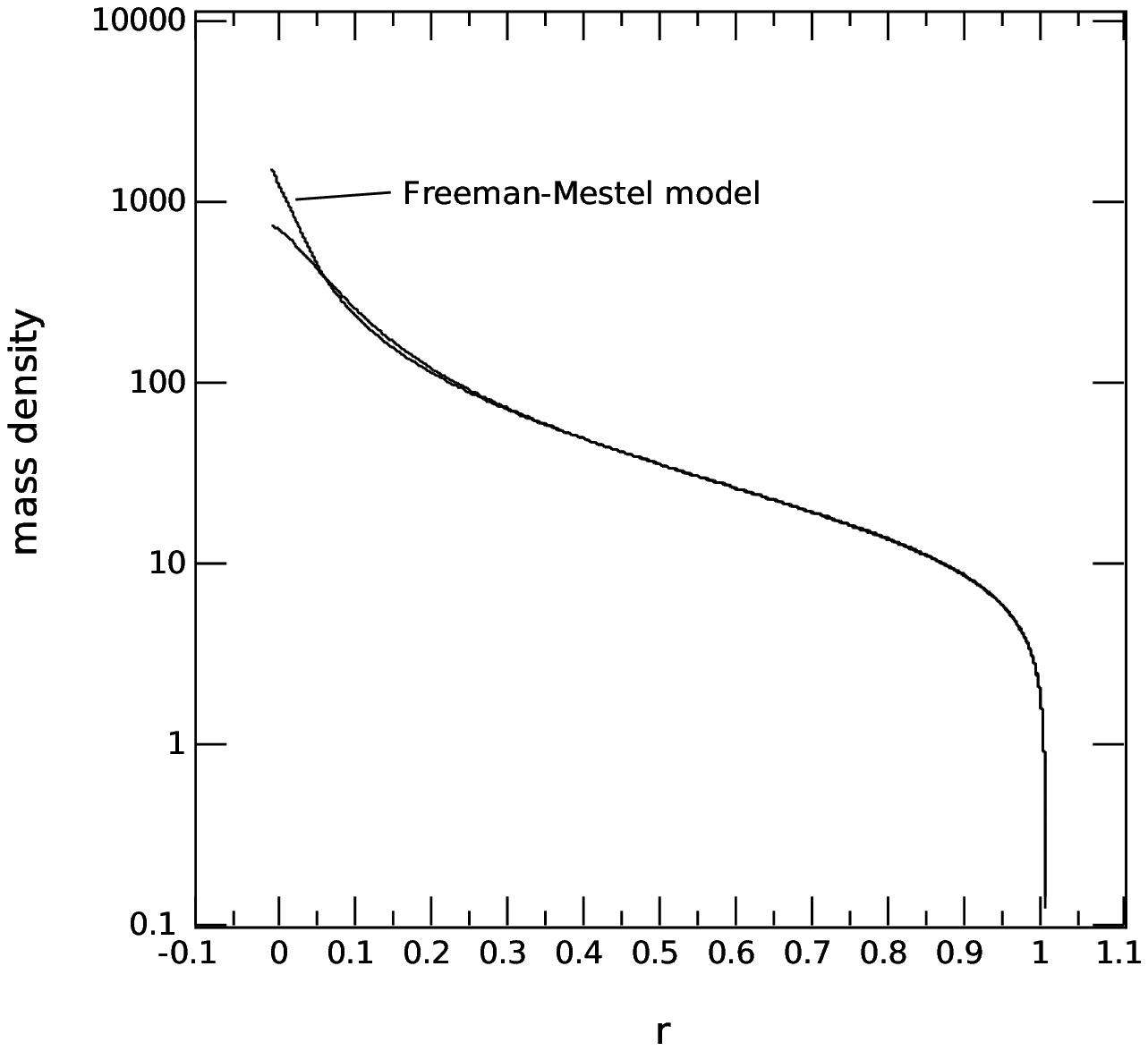}}
\caption{\label{fig:fig5}The distribution of
mass density $\rho(r)$ given by (25) for the
Freeman-Mestel model at $\tilde{R}_c = 0.05$ with $A = 1.6060$,
compared with that in Fig~\ref{fig:fig3} for 
$R_c = 0.05$ with $A = 1.5777$.}
\end{figure}

Although $\tilde{R}_c$ here also serves as a scaling parameter for the
galactic core, having a similar physical meaning as $R_c$ 
in (\ref{eq:V-ideal}), 
the value of $\tilde{R}_c$ does not have any mathematical relationship
with that of $R_c$. 
For example, at $\tilde{R}_c = 0.05$ (\ref{eq:freeman-mestel_rho})
and (\ref{eq:freeman-mestel_A}) yield 
$V(r)$ and $\rho(r)$ in Figs~\ref{fig:fig4} and \ref{fig:fig5} 
noticeably different from 
those in Figs~\ref{fig:fig2} and \ref{fig:fig3}. 
For smaller values of $\tilde{R}_c$, the differences between 
$\rho(r)$ given by the Freeman-Mestel model and that in
Fig~\ref{fig:fig3} at the same values of $R_c$ 
are less visually discernable. 
But the value of $A$ determined by the Freeman-Mestel model
can still be slightly different.
For example, at $\tilde{R}_c = R_c = 0.01$ (\ref{eq:freeman-mestel_A})
yields $A = 1.5777$ whereas that computed in \S~3.1 is $A = 1.5710$.
It seems for a given value of $\tilde{R}_c = R_c$ the rotation curve 
of the Freeman-Mestel model has a greater slope for the 
rising velocity in galactic core but a 
somewhat less flat velocity outside the core,
as shown in Fig~\ref{fig:fig4}.
Such a numerical difference tends to diminish with diminshing 
$\tilde{R}_c$, e.g., we have $A = 1.57147$, $1.57084$, and $1.57081$ 
for $\tilde{R}_c = 10^{-3}$, $10^{-4}$, and $10^{-5}$, respectively.
As expected, $A \to 1.57080$ as that for the Mestel disk given 
in (\ref{eq:A-Mestel}) at the limit of $\tilde{R}_c \to 0$. 

What we try to illustrate here is that 
for obtaining rotation curves with basic observed features,
a simple analytical mass density model as constructed by 
combination of those of \cite{mestel63} and \cite{freeman70}
in (\ref{eq:freeman-mestel_rho}) 
seems to be quite reasonable and convenient.
In terms of computational efforts, 
it is usually much easier and faster 
to compute the rotation velocity $V(r)$ from 
a given mass density distribution $\rho(r)$ than 
vice versa.  
This is because that computing $V(r)$ for a known $\rho(r)$ does 
not need to solve the matrix problem.
However, there has not been reliable means for directly measuring
the mass distribution in a disk galaxy.
The mass distribution derived from measured luminosity 
must rely on assumed mass-to-luminosity ratio, with the validity of which   
being a subject of debate. 
Thus, accurately measured rotation curves remain as the most reliable basis
for determining the distribution of mass in disk galaxies,
providing fundamental information for understanding 
the stellar dynamics in galactic disks \citep{sofue01}.

\subsection{Analysis of measured rotation curves of arbitrary shapes}

For rotation curves with ``idealized'' shapes expressed
in terms of simple mathematical functions like that in 
(\ref{eq:V-ideal}), we have shown that the 
numerically computed mass density distribution
$\rho(r)$ approaches that of Mestel's disk (\ref{eq:Mestel}) 
when the galactic core is small, e.g., for $R_c \le 0.02$.
But some measured rotation curves can vary significantly from
those described by simple mathematical functions 
or those produced by conveniently constructed mass density functions like 
with the Freeman-Mestel model (\ref{eq:freeman-mestel_rho}).

\begin{figure}%[htb]
\resizebox{!}{0.72\textwidth}
{\includegraphics[clip=true,scale=0.82,viewport=60 330 760 760]{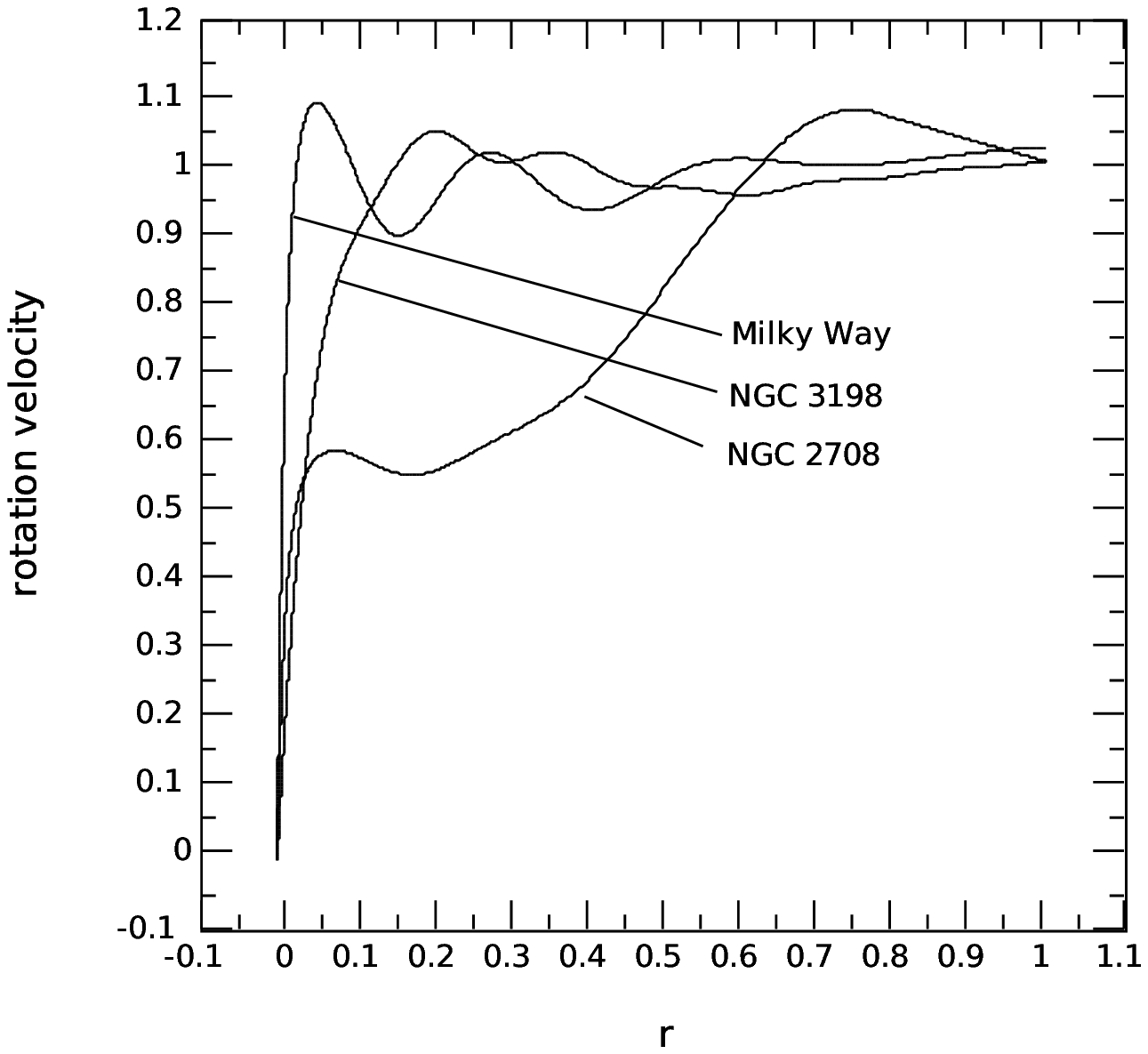}}
\caption{\label{fig:fig6}The rotation curves $V(r)$ 
of Milky Way with $V_0 = 2.2 \times 10^5$ (m s$^{-1}$) 
and $R_g = 4.73 \times 10^{20}$ (m),
NGC 3198 with $V_0 = 1.5 \times 10^5$ (m s$^{-1}$)   
and $R_g = 9.24 \times 10^{20}$ (m),
and NGC 2708 with $V_0 = 2.3 \times 10^5$ (m s$^{-1}$)   
and $R_g = 1.42 \times 10^{20}$ (m).}
\end{figure}

\begin{figure}%[htb]
\resizebox{!}{0.72\textwidth}
{\includegraphics[clip=true,scale=0.82,viewport=60 330 760 760]{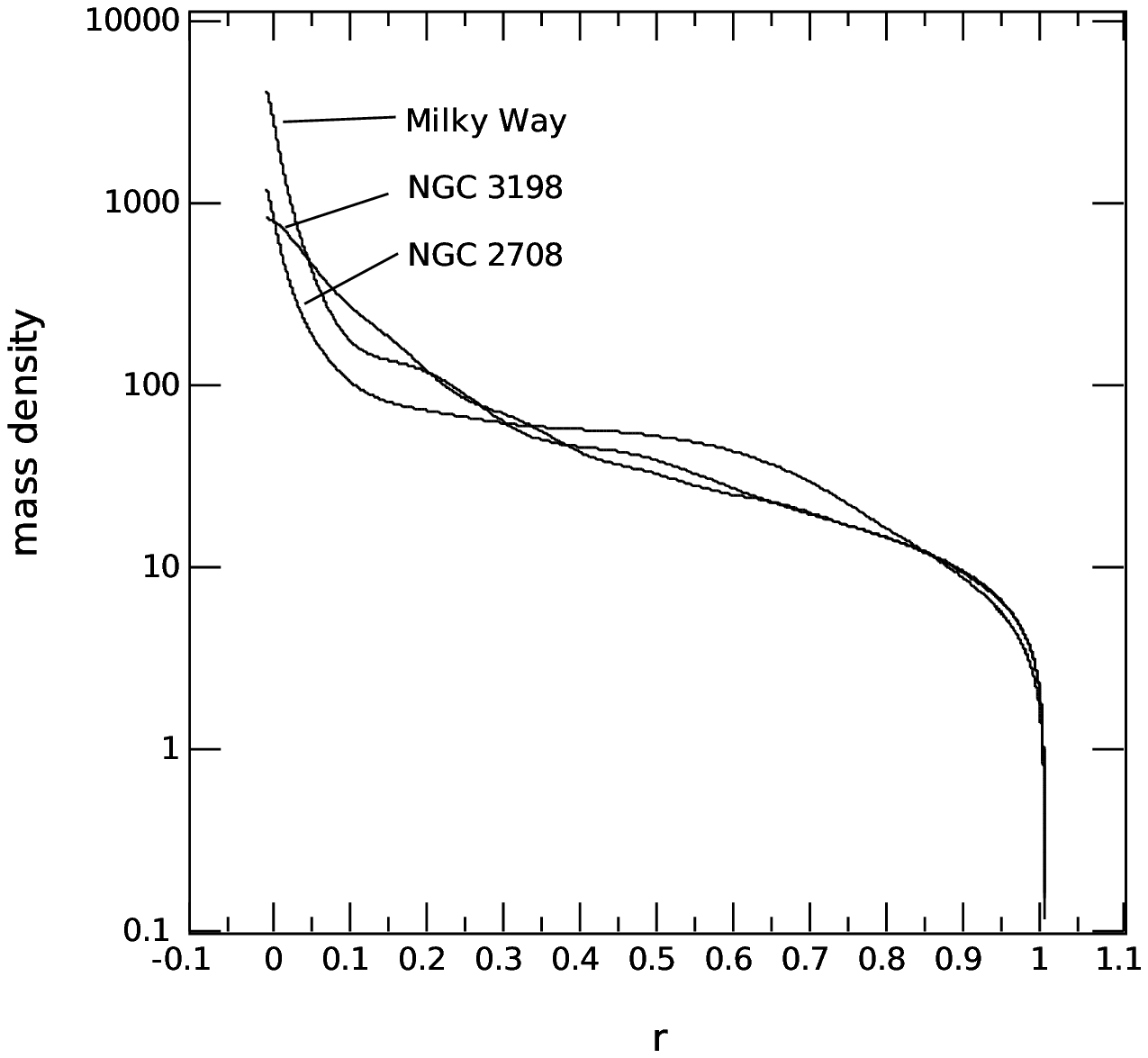}}
\caption{\label{fig:fig7}The computed 
mass density distributions $\rho(r)$ from given rotation curves 
in Fig~\ref{fig:fig6} for
Milky Way, NGC 3198, and NGC 2708, 
with the values of $A$ determined as $1.564$, $1.619$, 
and $1.644$, respectively.}
\end{figure}

To determine the mass density distribution according to Newtonian dynamics
from a measured rotation curve of arbitrary shape,
our computational scheme based on sound mathematical foundation as
presented in \S~2 (as well as Appendix A) 
can become a generally applicable and flexible tool 
for many practical applications. 
As an example, 
here in Figs~\ref{fig:fig6} and \ref{fig:fig7} 
we show our computed mass density distributions for 
a few actually measured galactic rotation curves 
with noticeably different characteristics.

The measured rotation curve for the Milky Way in Fig~\ref{fig:fig6}
seems to be just a few bumps and wiggles superposed on
that in Fig~\ref{fig:fig2} for $R_c = 0.01$.
Therefore, it is no surprise to 
see that the corresponding mass density curve for the Milky Way
in Fig~\ref{fig:fig7} also exhibits a few bumps and wiggles around  
that in Fig~\ref{fig:fig3} for $R_c = 0.01$. 
Similarly, the measured rotation curve for NGC 3198 in Fig~\ref{fig:fig6}
appears to be just 
that in Fig~\ref{fig:fig2} for $R_c = 0.05$ with 
some small perturbations,
and so does the computed NGC 3198 mass density in Fig~\ref{fig:fig7}
compared with that for $R_c = 0.05$ in Fig~\ref{fig:fig3}.
But the rotation curve for NGC 2708 in Fig~\ref{fig:fig6}
differs significantly from those of typical shapes in Fig~\ref{fig:fig2}.
The computed mass density distribution for NGC 2708 in Fig~\ref{fig:fig7}
shows noticeably different features from those 
in Fig~\ref{fig:fig3}. 
The sharp rise of mass density forward galactic center 
corresponds to a fast dropping of rotation velocity,
as required for the force balance in Newtonian dynamics. 
The gradual increase in the rotation velocity 
in the middle section $(0.1, 0.7)$
of NGC 2708 leads to a slow decreasing 
local mass density.
Then a slight reduction of the rotation velocity toward 
the galactic periphery is responsible for a faster decrease of 
local mass density in the outer region $r > 0.7$
than those for flat rotation curves  
in Fig~\ref{fig:fig7} for NGC 2708.

Despite the differences in rotation curves in Fig~\ref{fig:fig6},
the computed values of galactic rotation number $A$ for
these three galaxies are quite close within a few percents,
namely, $A = 1.564$, $1.619$, and $1.644$, respectively for
the Milky Way, NGC 3198, and NGC 2708.
This is consistent with that shown in Fig~\ref{fig:fig3}
for a wide range of $R_c$.
Thus, we may reasonably conclude that for most disk galaxies, the 
value of galatic rotation number is expected to be 
within $\pm 10\%$ of $A = 1.70$, with smaller $A$ for
the galaxies having high-density core and small $R_c$ 
and larger $A$ for those having more gradual rise in the rotation curve
with larger $R_c$.

Although we only computed examples with a few representative 
rotation curves, such as those described by the idealized formula 
(\ref{eq:V-ideal}) with several values of $R_c$ 
and those actually measured with different characteristics,  
we believe the cases examined here actually 
cover a wide enough range of observational measurements that 
our results can offer general physical insights.
Cases with rotation curves falling either close to or
in between those illustrated here are not expected to 
differ considerably from
our present findings. 

\section{Discussion}
The problem of determining the mass distribution
in a thin axisymmetric disk from observed circular velocities 
has been investigated by many authors over the past fifty years,
through various mathematical approaches.
Yet satisfactory method for accurate computation is still lacking,
despite the galactic rotation model has been simplified as much as possible 
for concisely describing only the most essential features. 
The main obstacle here appears to have been 
due to the mathematical singularities 
in the elliptic integrals that are apparently difficult to handle.
Here in this work,
we present an efficient, robust computational method 
with appropriate mathematical treatments such that  
the apparent difficulties associated with the singularities 
are completely removed.
Thus, we are enabled to systematically analyze 
the basic features in a rotating disk galaxy,
with properly nondimentionalized mathematical formulations.
Further refinement of the present galactic rotation model 
may provide description of some of the fine details 
such as the spiral arm structure with 
non-axisymmetic motion \citep{koda02}, 
gas pressure effect
in the central core \citep{dalcanton10}, 
disk thickness effect \citep{casertano83},
etc.
But those fine details should not alter 
the basic features significantly, 
at least in a gross qualitative sense.
Our results in Fig 7 show that 
the general shape of the mass density distribution 
remains quite similar for rotation curves of
drastically different appearances.
The value of the galactic rotation number $A$ 
does not change more than $\pm 10\%$ for a variety of 
rotation curves, 
indicating that the gross balance between 
the centrifugal force and gravitational force in a disk galaxy 
is usually insensitive to fine details.

\subsection{Total mass in galactic disk}

In the dimensionless form as presented here,
our mathematical system contains a dimensionless parameter
called galactic rotation number $A$.  
This galactice rotation number,
with its value determined as part of the computational solution,
can provide a unique insight into the dynamical system of 
rotating galaxy.
From the knowledge of $V_0$ and $R_g$ from measured
rotation curves, we can determine the value of total mass $M_g$
based on computed value of $A$
(cf. (\ref{eq:parameter-A})) as
\begin{equation} \label{eq:M_g}
M_g
= \frac{V_0^2 \, R_g}{A \, G} \, .
\end{equation}

According to the rotation curve of the Milky Way in Fig~6,
we have the galactic rotation number $A = 1.564$.
Then, from the measured values $V_0 = 2.2 \times 10^5$ (m s$^{-1}$)
and $R_g = 5 \times 10^4$ (light-years) $= 4.73 \times 10^{20}$ (m)
(which is about $15.3$ kpc where $1$ kpc $= 3.086 \times 10^{19}$ m),
(\ref{eq:M_g}) yields
\begin{equation} \label{eq:M_g-Milky}
M_g
= 2.19 \times 10^{41} (\mbox{kg}) = 1.10 \times 10^{11} (\mbox{solar-mass}) \, .
\nonumber
\end{equation}
(Here, $1$ solar-mass $= 1.98892 \times 10^{30}$ kg.)
This value is in very good agreement with the Milky Way star counts
of 100 billion
\citep{sparke07}.

Another example in Figs~6 and 7 is the galaxy NGC 3198,
with $V_0 = 1.5 \times 10^5$ (m s$^{-1}$) and
$R_g = 30$ (kpc) $= 9.24 \times 10^{20}$ (m) \citep{begeman8789}.
Using the computed $A = 1.619$, we obtain $M_g = 1.925 \times 10^{41}$ (kg)
$= 9.68 \times 10^{10}$ (solar-mass).

For a small disk galaxy NGC 6822,
we have a rotation curve similar to that described by $R_c \sim 0.3$
in (\ref{eq:V-ideal}),
with $V_0 = 6.0 \times 10^4$ (m s$^{-1}$) 
and $R_g = 5$ (kpc) $= 1.54 \times 10^{20}$ (m) \citep{weldrake03}.
If we take $A = 1.92$ for $R_c = 0.3$, 
(\ref{eq:M_g}) yields 
$M_g = 4.33 \times 10^{39}$ (kg)
$= 2.18 \times 10^{9}$ (solar-mass).

Because the value of $A$ does not vary much for a large range of 
rotation curves with various shapes 
(see, e.g., Figs~\ref{fig:fig6} and \ref{fig:fig7}), 
what (\ref{eq:M_g}) implies is
that $M_g \propto V_0^2 \, R_g$ 
as what \cite{bosma78} found from evaluating 
mass versus size in a large number of observed disk galaxies. 
For a fixed value of $V_0$, 
$M_g \propto R_g$. 
Therefore, a disk galaxy cannot physically extend 
indefinitely in size,
for $M_g$ to remain finite.
In other words, there must be an edge of the galactic disk at 
a finite radius $R_g$, 
where the mass density precipitously diminishes.
Normally, one would define
$R_g$ as the radial distance 
where the ``luminous'', ``visible'', or ``detectable'' signal for 
rotating matter ends.
With the advance in measurement technology using different emission lines,
the detectable rotating matter (in the form of gas) 
seems to extend further out from
the optically visible disk \citep[cf.][]{sofue01}.
Thus, the value of $R_g$ may change with the evolving 
astronomical observation technology.
Wherever the true $R_g$ is located, 
it must correspond to an abruptly steep decrease of mass density 
whereas the mass density variation within $R_g$ is expected to be 
smooth,
according to our Newtonian dynamics model for 
thin-disk galaxies with typical rotation curves.  
It should be noted that 
although for a given rotation curve with fixed $V_0$ the total mass $M_g$
of the galactic disk increases linearly with $R_g$,
the dimensional value of surface mass density 
should generally
decrease with $R_g$ according to $1/R_g$ 
because it scales as $M_g / R_g^2$.

As an interesting exercise,
we may take (\ref{eq:Mestel}) for the convenience in estimating the 
surface mass density $\sigma(r) \equiv \rho(r)\,h$ 
around the Sun in the Milky Way when $R_g$ increases.
Then,
we obtain $\sigma(r_{sun}) = 0.3106$, $0.7954$, and $1.7532$ 
for $r_{sun} = 0.5229$, $0.2614$, and $0.1307$, respectively 
for $R_g = 15.3$, $30.6$, and $61.2$ (kpc) assuming the Sun is 
located at $r_{sun}\,R_g = 8$ (kpc) from the galactic center.
Based on the value given by (\ref{eq:M_g-Milky}), 
we have the dimensional surface mass density 
$\sigma(r_{sun})\,M_g /R_g^2 \approx 146$ (solar-mass pc$^{-2}$)
for $R_g = 15.3$ (kpc).
If $R_g$ for the flat rotation curve 
were found to be at $30.6$ or $61.2$ (kpc), 
the dimensional surface mass density would become 
$187$ or
$206$ (solar-mass pc$^{-2}$), 
varying much less dramatically than the value of $R_g$.
This phenomenon is a consequence of 
the $1/r$ part of (\ref{eq:Mestel}), which becomes more dominant 
for smaller values of $r$.
In fact, if the surface mass density $\sigma(r)$ 
were strictly to follow a 
distribution $\propto 1/r$, 
the dimensional surface mass density for a given dimensional 
radial coordinate $r R_g$ would remain constant 
because the value of $A$ changes little if at all.
Thus, as $R_g$ extends further out, the value of 
dimensional surface mass density in the neighborhood of Sun
is expected to become almost independent of the value of $R_g$. 

\subsection{Computed mass density versus observed
surface brightness}

Observations of disk galaxies suggest that the 
surface brightness--the total stellar luminosity 
emitted per unit area of the disk--is approximately 
an exponential function of radius \citep{freeman70, binney87}.
This exponential approximation seems to be especially 
good for the outer part of disk galaxies where 
the inner bulge component diminishes \citep[e.g.,][]{freeman70}.
Our computed mass density distributions in Fig~\ref{fig:fig3} 
according to typical flat rotation curves 
indeed show nearly straight-line shape in the log-linear plots 
corresponding to approximately exponential function
for a large portion of the problem domain,
e.g., in the interval $(0.2, 0.9)$. 
In fact, the least-square fit of our computed $\ln\,\rho$ versus $r$
for the case of $R_c = 0.01$ (cf. Fig~\ref{fig:fig3})
to a linear function
for $0.2 \le r \le 0.9$ yields
\begin{equation} \label{eq:least-square}
\ln\,\rho
= 5.2614 - 3.4377 \, r  \, ,
\end{equation}
with the correlation coefficient ``$R^2$'' being $0.9968$
suggesting that the portion of mass density
in $(0.2, 0.9)$ can indeed be well described 
by an exponential function $\rho = \rho_0 \, e^{-r/R_d}$ 
with $\rho_0 = 192.75$ and $R_d = 0.2909$. 
If the same least-square fitting were done 
for $0.1 \le r \le 0.9$, we would have 
$\rho_0 = 238.41$ and $R_d = 0.2668$ but with 
a slightly reduced correlation coefficient $R^2 = 0.9870$,
which still indicates a good approximation with 
the exponential function.
However, the dimensional ``radial scale length'' $R_d\,R_g$ for the 
Milky Way would be $\sim 4.5$ (or $4.1$) (kpc) according to 
$R_d = 0.2909$ (or $0.2668$) assuming $R_g = 15.3$ (kpc).
This is larger than the radial scale length 
$2.5$ (kpc) from fitting the brightness 
measurement data reported by \cite{freudenreich98}. 
For NGC 3198 with $R_g = 30$ (kpc), we would have $R_d\,R_g = 8.73$ 
(or $8.00$) (kpc), again larger than the radial scale length of $2.63$ (kpc) 
for the luminosity profile \citep[cf.][]{begeman8789}.
So, our computed results
suggest that the surface mass density decreases toward the galactic
periphery at a slower rate than that of the luminosity density.
In other words, the mass-to-light 
ratio in a disk galaxy is not a constant; 
it generally increases with the radial distance 
from the galactic center 
as indicated by 
our analysis for the exponential portion of mass density distribution
\citep[which was also suggested by][]{bosma78}. 

But it is known that the constructed  mass density distribution 
in terms of a single exponential function 
cannot generate an observed flat rotation curve \citep{freeman70, binney87}.
The sharp increase of the mass density near the galactic center
that drastically deviates the exponential description for 
$0.1 \le r \le 0.9$
or
$0.2 \le r \le 0.9$
seems to play an important role for keeping the rotation curve
flat forward the galactic center up to the edge of the core.
In reality, most disk galaxies also have a central bulge 
with apparently high concentration of stars. 
Our pure disk model does not explicitly treat the bulge as 
a separate object; instead, the gravitational effect of the bulge 
is lumped in the rotating disk.
Thus, our computed mass density should be regarded as 
a combination of that from the pure disk 
and the effective bulge represented in the disk form.
This sharp increase of the disk mass density near the galactic center
can be considered as an account for the highly concentrated 
mass in the central bulge.  
Actually, it may not be impossible to extend the formulation in \S~3.2
for a mass density distribution to include a summation (or expansion)
of several exponential terms with different radial scales lengths, 
for matching an observed rotation curve with more complicated shape.
Yet, the most straightforward method 
for determining the mass density distribution
for a given rotation curve (of arbitrary shape) is 
by numerically solving the linear algebra matrix equation 
derived based on sound mathematical ground for disk galaxies of finite size
as presented in \S~2 and demonstrated in \S~3.1 and \S~3.3.

\subsection{Inaccuracy of Keplerian dynamics
for disk galaxies}

Enchanted by its simplicity, 
the Keplerian dynamics was applied by several authors
in description of the disk galaxy behavior
without seriously inquiring its validity and accuracy.
To clarify some of the problems in such an over-simplification,
here we present a quantitative analysis of the fundamental differences 
between the Keplerian dynamics and Newtonian dynamics 
especially when applied to disk galaxies.
 
From analyzing the orbits of planets around the Sun,
Kepler empirically discovered laws for planet motion in the solar system.
It was Newton who mathematically showed that Kepler's laws are
actually consequences of Newton's laws of motion and universcal law
of gravitation.
The so-called Keplerian dynamics can be derived from Newton's theorems for
the gravitational potential of any spherically symmetric
mass distribution.
In considering the balance between gravitational force from
the distributed mass in a galaxy and centrifugal force due to rotation,
applying Keplerian dynamics would lead to an equation as
\begin{equation} \label{eq:force-balance-Keplerian}
\frac{2 \pi}{r^2} \int_0^r \rho(\hat{r})\,h\,\hat{r}d\hat{r}
- A\,\frac{V(r)^2}{r} = 0   \, ,
\end{equation}
which is apparently quite different from (\ref{eq:force-balance0})
as rigorously derived for the thin-disk galaxies.
However, the force balance equation based on Keplerian dynamics
(\ref{eq:force-balance-Keplerian}) looks much simpler
than that of Newtonian dynamics (\ref{eq:force-balance0}).
If justifiable in a quantitative sense, it may be
conveniently used as a reasonable approximation to the more involved
rigorous computations.
To provide a quantitative comparison,
we herewith examine a few basic mathematical features
of (\ref{eq:force-balance-Keplerian})
to illustrate whether the Keplerian dynamics can be
practically used as a reasonable approximation to
the Newtonian dynamics (\ref{eq:force-balance0}) for disk galaxies.

For a given rotation curve with the orbital velocity $V(r)$
described by (\ref{eq:V-ideal}),
an analytical solution to (\ref{eq:force-balance-Keplerian})
for $\rho(r)$ can be obtained as
\begin{eqnarray} \label{eq:rho-Keplerian}
\rho(r) = \frac{A}{2 \pi\,h} \left[\frac1r
\left(1 - 2 e^{-r/R_c} + e^{-2 r/R_c}\right) %\right. \nonumber \\
%\left.
+ \frac{2}{R_c}\left(e^{-r/R_c} - e^{-2r/R_c}\right)\right] \, .
\end{eqnarray}
Thus, (\ref{eq:rho-Keplerian}) describes a mass density
approaching $3A\,r/(2\pi\,h\,R_c^2) \to 0$ as $r \to 0$ with a positive
slope for small $r$ yet approaching $A/(2\pi\,h\,r)$ as $r \to 1$
(because $e^{-1/R_c}$ can be negligibly small for small $R_c$,
e.g., $e^{-1/R_c} = 4.54 \time 10^{-5}$, $2.06 \times 10^{-9}$,
and $1.93 \times 10^{-22}$ for $R_c = 0.1$, $0.05$, and $0.02$,
respectively).
The mass density distribution of (\ref{eq:rho-Keplerian}) does
not monotonically decrease with $r$ as that shown in Fig~{\ref{fig:fig3};
instead, it is zero at the galactic center and increases for small $r$
according to a slope $\propto 1/R_c^2$ (which can be large for small $R_c$)
until reaching a peak value,
then decreases in a form $\propto 1/r$ towards the galactic periphery
$r = 1$ without the precipitous drop seen in Fig~\ref{fig:fig3}.

Substituting (\ref{eq:rho-Keplerian}) to (\ref{eq:mass-conservation})
yields
\begin{equation} \label{eq:A-Keplerian}
A = \frac{1}{1 - 2 e^{-r/R_c} + e^{-2 r/R_c}} \, ,
\end{equation}
which leads to $A \approx 1$ for small $R_c$,
quite different from $1.57$ when $R_c \to 0$ as obtained in \S~3.1.
Hence using the Keplerian dynamics to describe the disk galaxies
can be misleading, because not only the results differ quantitatively
but also qualitatively from that based on rigorous computations.

On the other hand, if we assume the mass density distribution is
known, e.g., as that given by (\ref{eq:Mestel}),
(\ref{eq:force-balance-Keplerian}) leads to
\begin{eqnarray} \label{eq:V-Keplerian}
V(r)^2 = \frac{1}{r}\int_0^r\left[1 - \frac{2}{\pi} \sin^{-1}(\hat{r})\right]
d\hat{r} \nonumber %\\
= 1 - \frac{2}{\pi} \left[\sin^{-1}(r)
- \frac{1 - \sqrt{1 - r^2}}{r}\right] \, .
\end{eqnarray}
Instead of a completely flat rotation curve,
the Mestel's disk mass density distribution with Keplerian dynamics
would yield orbital velocity $V(r)$ that monotonically decreases with $r$,
having $V(0) = 1$ and $V(1) = 0.7979$.
Therefore, a mass density distribution
corresponding to a flat rotation curve based on Newtonian dynamics
would be mistaken as failing to explain the observed flat rotation curve
when Keplerian dynamics
were inappropriately employed,
because it instead predicts a falling rotation curve.

\section{Conclusions}

In the present paper, we show that
with appropriate mathematical treatments 
the apparent difficulties associated with singularities in 
computing elliptic integrals can be eliminated 
when modeling Newtonian dynamics of thin-disk galactic rotation.
Using the well-established boundary element techniques,
the nondimensionalized governing equations for disks of finite sizes 
can be discretized, transformed into a linear algebra matrix equation,
and solved by straightforward Gauss elimination
in one step without further iterations. 
Although the mathematical derivations in Appendix A 
for removing the singularities seem somewhat sophisticated,
the actual implementations of the numerical code are not as lengthy.
With our code on a typical personal computer with a single Pentium 4 processor, 
each solution in \S~3 takes no more than a minute or so to compute. 
Thus, a numerical code implemented according to our algorithm 
can be conveniently used to accurately determine 
the surface mass density distribution in a disk galaxy 
from a measured rotation curve (or vice versa),
which is important for in-depth understanding of the 
Newtonian dynamics and its capability of 
explaining the ``galaxy rotation problem'' via rotation curve analysis.
Moreover, the dimensionless galactic rotation number $A$
in our mathematical system can provide important insights 
into the general dynamical behavior of disk galaxies.  

Through a systematic computational analysis, we have illustrated that
the value of the galactic rotation number remains within 
$\pm 10\%$ of $A = 1.70$ for a wide variety of 
rotation curves. 
For most Sb type galaxies like the Milky Way,
having rotation curves typically with a very steep rise in
a small central core region and a large range of flat portion,
we have showed that $A \approx 1.60$ with a surface mass density 
very close to that of Mestel's disk 
(except in an infinitestmal neighborhood of the galactic center
where the Mestel disk becomes singular).  
But for galaxies with ``non-ideal'' rotation curves containing considerable 
irregularities, 
our numerical approach can easily be used without modification for 
computing the corresponding surface mass density distributions accurately
for rotation curve analysis. 

Because the value of $A \equiv V_0^2\, R_g / (M_g\, G)$ 
remains almost invariant for various galaxies,
we can draw a conclusion that the total mass in a disk galaxy
$M_g$ must be proportional to $V_0^2 \, R_g$.
For galaxies with similar characteristic rotation velocity $V_0$,
their total mass $M_g$ must be proportional to their disk size $R_g$.
Our model predicts that at the disk edge 
the surface mass density is expected to diminish precipitously 
whereas within the disk edge the surface mass density should 
vary rather smoothly without sharp changes except near the galactic center.
Thus, a disk galaxy with a finite amount of mass must also have a 
finite size, based on the Newtonian dynamics modeling.

For a disk galaxy with a typical flat rotation curve,
our modeling result show that the surface mass density 
monotonically decreases from the galactic center toward periphery,
according to Newtonian dynamics.
In a large portion of the galaxy, the surface mass density 
follows an approximately exponential law of decay 
with respect to the galactic radial coordinate.
Yet the radial scale length for the exponential portion of 
surface mass density seems to be generally larger than that of 
the measured exponential brightness distribution,
suggesting an increasing mass-to-light ratio 
with the radial distance in a disk galaxy.
This is consistent with 
typical edge-on views of disk galaxies often revealing 
a dark edge against a bright background bulge.

\normalem
%\begin{Acknowledgements}
\section*{Acknowledgements}
We are indebted to Dr. Len Gray of Oak Ridge National Laboratory 
for teaching detailed boundary element techniques 
for elegant removal of various singularities in integral equations.
We want to thank Dr. Louis Marmet for his intuitive discussion
and preliminary computational results 
that convinced us to pursue a rigorous numerical analysis 
of the galactic rotation problem. 
The results shared by Ken Nicholson, Prof. Michel Mizony 
for computing the similar problem should also be acknowledged 
for enhancing our confidence.
%\end{Acknowledgements}

\appendix

\section{Treatments of Singular Elements}
The complete elliptic integrals of the first kind and second kind can
be numerically computed with the formulas \citep{abramowitz72}
\begin{equation} \label{eq:K-m1}
K(m) = \sum_{l = 0}^4 a_l m_1^l - \log(m_1) \sum_{l = 0}^4 b_l m_1^l
\end{equation}
and
\begin{equation} \label{eq:E-m1}
E(m) = 1 + \sum_{l = 1}^4 c_l m_1^l - \log(m_1) \sum_{l = 1}^4 d_l m_1^l \, ,
\end{equation}
where
\begin{equation} \label{eq:m1-def}
m_1 \equiv 1 - m = \left(\frac{\hat{r} - r}{\hat{r} + r}\right)^2 \, .
\end{equation}
Clearly, the terms associated with $K(m_i)$ and $E(m_i)$ in
(\ref{eq:force-balance-residual}) become singular when $\hat{r} \to r_i$
on the elements with $r_i$ as one of their end points.

The logarithmic singularity can be treated by converting the
singular one-dimensional integrals into non-singular
two-dimensional integrals
by virtue of the identities:
\begin{eqnarray} \label{eq:log-integral-identities}
\left\{
\begin{array}{cc}
\int_0^1 f(\xi) \log \xi d\xi = - \int_0^1 \int_0^1 f(\xi \eta) d\eta d\xi
\\
\int_0^1 f(\xi) \log(1 - \xi) d\xi =
- \int_0^1 \int_0^1 f(1 - \xi \eta) d\eta d\xi  \,
\end{array}  \right . \, ,
\end{eqnarray}
where $f(\xi)$ denotes a well-behaving (non-singular) function of $\xi$
on $0 \le \xi \le 1$.

But a more serious non-integrable
singularity $1 / (\hat{r} - r_i)$ exists due to
the term $E(m_i) / (\hat{r} - r_i)$ in
(\ref{eq:force-balance-residual}) as $\hat{r} \to r_i$.
The $1 / (\hat{r} - r_i)$ type of singularity is treated by
taking the Cauchy principle value to 
obtain meaningful evaluation \citep[cf.][]{kanwal96},
as commonly done with the boundary element method 
\citep{sladek98, gray98,sutradhar08}.
In view of the fact that each $r_i$ is considered to be shared by two
adjacent elements covering the intervals $[r_{i-1}, r_i]$ and
$[r_i, r_{i+1}]$, the Cauchy principle value of
the integral over these two elements is given by
\begin{equation} \label{eq:CPV-def}
\lim_{\epsilon \to 0} \left[
\int_{r_{i-1}}^{r_i -\epsilon} \frac{\rho(\hat{r}) \hat{r} d\hat{r}}{
\hat{r} - r_i}
+ \int_{r_i + \epsilon}^{r_{i+1}} \frac{\rho(\hat{r}) \hat{r} d\hat{r}}{
\hat{r} - r_i}\right]
\, .
\end{equation}
In terms of elemental $\xi$, (\ref{eq:CPV-def}) is equivalent to
\begin{eqnarray} \label{eq:CPV-xi}
-\lim_{\epsilon \to 0} \left\{
\int_0^{1 -\epsilon/(r_i - r_{i-1})} \frac{[\rho_{i-1} (1 - \xi) + \rho_i \xi]
[r_{i-1} (1 - \xi) + r_i \xi] d\xi}{
1 - \xi} \right . \nonumber\\
\left .
-\int_{\epsilon/(r_{i+1}-r_i)}^1 \frac{[\rho_{i} (1 - \xi) + \rho_{i+1} \xi]
[r_i (1 - \xi) + r_{i+1} \xi] d\xi}{
\xi} \right\}
\, .
\end{eqnarray}
Performing integration by parts on (\ref{eq:CPV-xi}) yields
%\begin{widetext}
\begin{eqnarray}
\rho_i \, r_i \log\left(\frac{r_{i+1}-r_i}{r_i-r_{i-1}}\right) 
-\left(
\int_0^1 \frac{d\{[\rho_{i-1} (1 - \xi) + \rho_i \xi]
[r_{i-1} (1 - \xi) + r_i \xi]\}}{d\xi} 
\log(1 - \xi) d\xi \right . \nonumber\\
\left .
+\int_0^1 \frac{d\{[\rho_{i} (1 - \xi) + \rho_{i+1} \xi]
[r_i (1 - \xi) + r_{i+1} \xi]\}}{d\xi} \log \xi d\xi
\right)
\, , \nonumber
\end{eqnarray}
%\end{widetext}
where the two terms associated with $\log \epsilon$ 
cancel out each other, the terms with $\epsilon \log \epsilon$ 
become zero at the limit of
$\epsilon \to 0$, and the first term becomes nonzero when the mesh nodes
are not uniformly distributed (namely, the adjacent elements are 
not of the same segment size).  In other words, inclusion of this first term 
enables the usage of nonuniformly distributed nodes for more effective
computations, which is one of the algorithm improvements 
over that in our previous works 
\citep{gallo09, gallo10}. 

At the galaxy center $r_i = 0$ (i.e., $i = 1$),
\begin{equation} \label{eq:at_r=0}
\int_{r_i}^{r_{i + 1}} \frac{\rho(\hat{r}) \hat{r} d\hat{r}}{
\hat{r} - r_i} = \int_0^{r_{i + 1}} \rho(\hat{r}) d\hat{r} \, .
\end{equation}
Thus, the $1/(\hat{r} - r_i)$ type of singularity disappears naturally.
However, numerical difficulty can still arise 
if $\rho$ itself becomes singular
as $r \to 0$, e.g., $\rho \propto 1/r$ 
as for the Mestel disk \citep{mestel63}.
The singular mass density at $r = 0$ corresponds to a mathematical cusp,
which usually indicates the need of finer resolution in the physical space. 
To avoid the cusp in mass density at the galactic center,
we can impose a requirement of 
continuity of the derivative of $\rho$ at the galaxy center $r = 0$.
This be easily implemented at the first node $i = 1$
to demand $d\rho / dr = 0$ at $r = 0$. 
In discretized form for $r_1 = 0$ we simply have
\begin{equation} \label{eq:rho-1}
\rho(r_1) = \rho(r_2)
\, .
\end{equation}

When $r_i = 1$ (i.e., $i = N$), we are at the end node 
of the problem domain.
Here we use a numerically relaxing boundary condition
by considering an additional element beyond
the domain boundary covering the interval
$[r_i, r_{i+1}]$, because it is needed to obtain a meaningful 
Cauchy principle value.  
In doing so we can also assume $r_{i+1} - r_i$ $= r_i - r_{i-1}$
such that $\log[(r_{i+1} - r_i)/(r_i-r_{i-1})]$ becomes zero,
to simplify the numerical implementation.
Moreover, it is reasonable to assume $\rho_{i+1} = 0$ because it 
is located outside the disk edge where the extremely low intergalactic 
mass density is expected to have inconsequential gravitational effect.
With sufficiently fine local discretization,
this extra element can be considered to
cover a diminishing physical space
such that its existence becomes
numerically inconsequential.
Thus, at $r_i = 1$ (where $i = N$) we have
\begin{eqnarray}
\int_0^1 \frac{d\{[\rho_{i} (1 - \xi) + \rho_{i+1} \xi]
[r_i (1 - \xi) + r_{i+1} \xi]\}}{d\xi} \log \xi d\xi \nonumber \\
=  (\rho_{i + 1} - \rho_i) \int_0^1 r(\xi) \log \xi d\xi
%\quad \quad \quad \quad  \nonumber \\
+ (r_{i+1} - r_i) \int_0^1 \rho(\xi) \log \xi d\xi %\nonumber \\
= \rho_i [r_i - \frac32(r_i - r_{i-1})] \, .
\nonumber
\end{eqnarray}
Now that only logarithmic singularities
are left, (\ref{eq:log-integral-identities}) can be used to eliminate
all singularities in computing the integrals
in (\ref{eq:force-balance-residual}).

Noteworthy here is that the (removable) singularities in the kernels of 
the integral equation (\ref{eq:force-balance}),
when properly treated, 
lead to a diagonally dominant Jacobian matrix 
with bounded condition number
in the Newton-Raphson formulation \citep{press88}.
This fact makes the matrix equation robust for any straightforward 
matrix solver.

\bibliographystyle{aa}

\label{lastpage}

\end{document}